\documentclass[twocolumn]{aastex701} 

\usepackage[T1]{fontenc}
\usepackage{ae,aecompl}
\usepackage{hyperref}
\usepackage{url}

\usepackage{longtable}
\usepackage{booktabs}
\usepackage{tabu}
\usepackage{graphicx}
\usepackage{multirow}
\usepackage{ulem}
\usepackage{color}
\usepackage{amsmath}

\def \s{~\rm{s}}

\def \K{~\rm{K}}

\def \g{~\rm{g}}
\def \G{~\rm{G}}

\def \yr{~\rm{yr}}
\def \Myr{~\rm{Myr}}
\def \Gyr{~\rm{Gyr}}

\usepackage{xcolor}
\definecolor{redak}{rgb}{0.9,0.15,0.05}

\shorttitle{Tycho is a SNIP}
\shortauthors{Soker}

\graphicspath{{./}{figures/}}

\begin{document}

\title{Tycho supernova exploded inside a planetary nebula (SNIP) }

\author[0000-0003-0375-8987]{Noam Soker} 
\affiliation{Department of Physics, Technion Israel Institute of Technology, Haifa, 3200003, Israel; soker@physics.technion.ac.il}
\email{soker@physics.technion.ac.il}


\begin{abstract}
I examine recent observations of the type Ia supernova remnant (SNR Ia) Tycho and conclude that Tycho is an SN Ia inside a planetary nebula (SNIP), strengthening such a previous suggestion from 1985. The observations reveal two opposite protrusions, termed ears, projected on the main shell of Tycho. The pair of ear structures resembles that of the SNRs Ia Kepler, SNR G299-2.9, and SNR G1.9+0.3, which earlier studies considered as SNIPs. 
The requirement that the explosion occurs within hundreds of thousands of years after the formation of the planetary nebula (by the second star to evolve) makes the core-degenerate scenario the most likely for Tycho. Several other possible scenarios lead to an SNIP, but they are unlikely for Tycho. The identification of Tycho as an SNIP leads to two general conclusions. (1) The fraction of SNIPs among normal SNe Ia is very large, $\approx 70-90\%$. Namely, the vast majority of normal SNe Ia are SNIPs. (2) To accommodate the large fraction of SNIPs, the delay time distribution of normal SNe Ia includes not only the stellar evolution timescale (as usually assumed), but also includes pockets of younger stellar populations in galaxies without ongoing star formation; the SNIPs come from the younger stellar populations in galaxies. 
\end{abstract}


\section{INTRODUCTION}
\label{sec:intro}

There is no consensus on the classification of type Ia supernova (SN Ia) scenarios as evident from the many reviews in the last decade (\citealt{Maozetal2014, MaedaTerada2016, Hoeflich2017, LivioMazzali2018, Soker2018Rev, Soker2019Rev, Soker2024Rev, Wang2018,  Jhaetal2019NatAs, RuizLapuente2019, Ruiter2020, Aleoetal2023, Liuetal2023Rev, Vinkoetal2023, RuiterSeitenzahl2025}). All scenarios, regardless of the classification, struggle to explain some observations, with some encountering severe difficulties. Some observations challenge most scenarios, or even all (e.g., \citealt{Pearsonetal2024, SchinasiLembergKushnir2025, sharonKushnirWygoda2025, Sharonetal2025, Wangetal2024}). 
There are scenarios with two or more channels (or sub-scenarios) of the main scenarios, e.g., the HeCO hybrid channel of the double degenerate (DD) scenario (e.g., \citealt{Peretsetal2019, Zenatietal2019, Zenatietal2023}), the common-envelope wind model \citep{MengPodsiadlowski2017, CuiMeng2022, WangMeng2025}, the channels of the single degenerate (SD) scenario with a main sequence star and a carbon–oxygen–neon white dwarf (WD) (e.g., \citealt{GuoMeng2025RAA}), and the core-merger detonation model \citep{Ablimit2021}. Some studies consider new processes, like new mass-transfer prescriptions (e.g., \citealt{Lietal2023RAA}). There is no one emerging leading scenario; therefore, studies in recent years have considered all scenarios and most of their channels for normal and peculiar SNe Ia (e.g., \citealt{Boraetal2024, Bossetal2024, Bregmanetal2024, CasabonaFisher2024, DerKacyetal2024, Joshietal2024, Koetal2024, Kobashietal2024, Limetal2024, Mandaletal2024, Mandaletal2025SNR0509,  MandalSetal2025Tycho, Mehtaetal2024, Palicioetal2024, Phillipsetal2024, Soker2024RAAPN, Uchidaetal2024, Burmesteretal2025, ChenCetal2025, Courtetal2025, Gabaetal2025NewA, Glanzetal2025a, Glanzetal2025b, Griffithetal2025, Hoogendametal2025a, Hoogendametal2025b, Itoetal2025, IwataMaeda2025, Kumaretal2025, Mageeetal2025, MichaelisPerets2025, OHoraetal2025, Panetal2025, Pollinetal2025, Simotasetal2025, WangChenPan2025, Bhatetal2026, Kwoketal2026, Pakmoretal2026}, for a limited list of papers since 2024). 

In addition to the general discussion of the scenarios, there are debates on specific SNe Ia, such as supernova remnant (SNR) 0509-67.5, where \cite{Dasetal2025NatAs} and \cite{Mandaletal2025SNR0509} argued for the double detonation (DDet) scenario; in contrast, I argued for the core-degenerate (CD) scenario \citep{Soker2025SNR0509}. In general, despite its popularity among many researchers (e.g., \citealt{Callanetal2024, MoranFraileetal2024, PadillaGonzalezetal2024, Polinetal2024, Shenetal2024, Zingaleetal2024, Rajaveletal2025, Wuetal2025}), the DDet scenario encounters challenges in explaining normal SNe Ia (e.g., \citealt{BraudoSoker2024, BraudoSoker2025RAA, Soker2024Comment, Soker2025SNR0509}) but might be more relevant to peculiar SNe Ia (e.g., \citealt{Glanzetal2026}) and in forming runaway WDs (e.g., \citealt{Glanzetal2025a}). Some runaway WDs might result from type Iax SNe (e.g., \citealt{Igoshevetal2023}), a group of peculiar SNe Ia, rather than the DDet scenario for normal SNe Ia.  

\cite{DickelJones1985} already suggested that Tycho exploded inside a planetary nebula. They performed hydrodynamic analytical calculations in spherical symmetry and concluded that a planetary nebula can account for both the thin outer shell and the main shell. In this study, I strengthen their suggestion by considering the non-spherical morphology of Tycho (Section \ref{sec:Geometry}). 
In Section \ref{sec:Scenarios}, I explain the CD, DD, and DDet scenarios in more detail and apply them to explain the properties of the Tycho SNR that I describe in Section \ref{sec:Geometry}. In Section \ref{sec:Implication} I discuss the implications of my claim that Tycho is a supernova Ia inside a planetary nebula (SNIP) on the fraction of SNIPs among all normal SNe Ia, and on the delay time distribution (DTD). 
I summarize this study in Section \ref{sec:Summary}.

\section{The two ears of Tycho SNR}
\label{sec:Geometry}

Tycho is a well-studied SN Ia, with hundreds of papers examining various aspects (e.g., \citealt{Ellienetal2023, Ferrazzolietal2023, Hinseetal2023, Petruketal2024, Picquenotetal2024, HanoverLeising2025, HollandAshfordetal2025}, for a small sample of recent papers; more in the text below), including its surroundings (e.g., \citealt{Ariasetal2019}). 

My starting point is recent studies that strongly indicate that Tycho is interacting with a circumstellar material (CSM).  
\cite{Uchidaetal2024} find rapid deceleration near the edge of Tycho SNR in almost every direction, inferring the presence of a dense wall surrounding Tycho; the inner volume is a wind bubble filled with a stellar progenitor wind (for a wind CSM, see also \citealt{Chiotellisetal2013}). \cite{Tanakaetal2021} find that a substantial deceleration started around the year $\simeq 2005$. \cite{Tanakaetal2021} and \cite{Uchidaetal2024} concluded that Tycho’s SNR is likely of SD origin; however, they do not mention the CD scenario at all, and, therefore, I think they reached a wrong conclusion (Section \ref{sec:Scenarios}). 
\cite{Kobashietal2024} and \cite{Kobashietal2025} study the interaction of the Tycho SNR with CSM. \cite{Kobashietal2024} estimate the CSM mass to be $M_{\rm CSM} \simeq 1.3-1.7 M_\odot$, depending also on the center of explosion. They find that a CSM formed by a stellar wind better fits their results than a constant-density CSM or ISM. They mention the possibility that the CD scenario might be the best scenario for Tycho. I go beyond that in arguing that Tycho is an SNIP, namely, an SN Ia inside a planetary nebula (or the remnant of a planetary nebula).   
\cite{Kobashietal2024} attribute the non-spherical expansion to non-spherical CSM, contrary to \cite{Millardetal2022}, who consider Tycho explosion to be non-spherical with higher explosion velocity towards the southeast.  
\cite{Kobashietal2024} find the mass loss rate that formed the CSM to be too high for the SD scenario, and prefer the CD scenario. \cite{Kobashietal2025}, in a later paper, write that the results support the SD scenario.  

The second point regarding Tycho, which became clear recently, is that despite its large-scale spherical structure on the plane of the sky, Tycho actually exhibits a significant departure from spherical symmetry.
\cite{PetrukKuzyo2025} studied the turbulence in Tycho and found it to be consistent with Kolmogorov-type turbulence. They found the X-ray correlation length of the two-point autocorrelation to have a long axis from northeast to southwest (at $35^\circ$ to north-south direction). \cite{Petruketal2025} find asymmetries in the silicon and sulfur distributions, which are not identical; the level of isotropy is higher in Si-rich plasma.  
\cite{Millardetal2022} find that the blueshifted knots are projected more in the northern region of Tycho, while redshifted knots are concentrated in the southern part of Tycho; other studies also find this large-scale asymmetry (e.g., \citealt{Godinaudetal2025}). 
\cite{Godinaudetal2023} performed the most detailed three-dimensional reconstruction. The general structure of Tycho they find is elongated, with an axis at a small angle to the line of sight. The small tilt of the axis is such that the approaching side (blueshifted) is towards the north and the receding (redshifted) is towards the south, as the above-cited papers found. 

\begin{figure*}[]
	\begin{center}
\includegraphics[trim=0.0cm 12.3cm 0.0cm 0.0cm ,clip, scale=0.82]{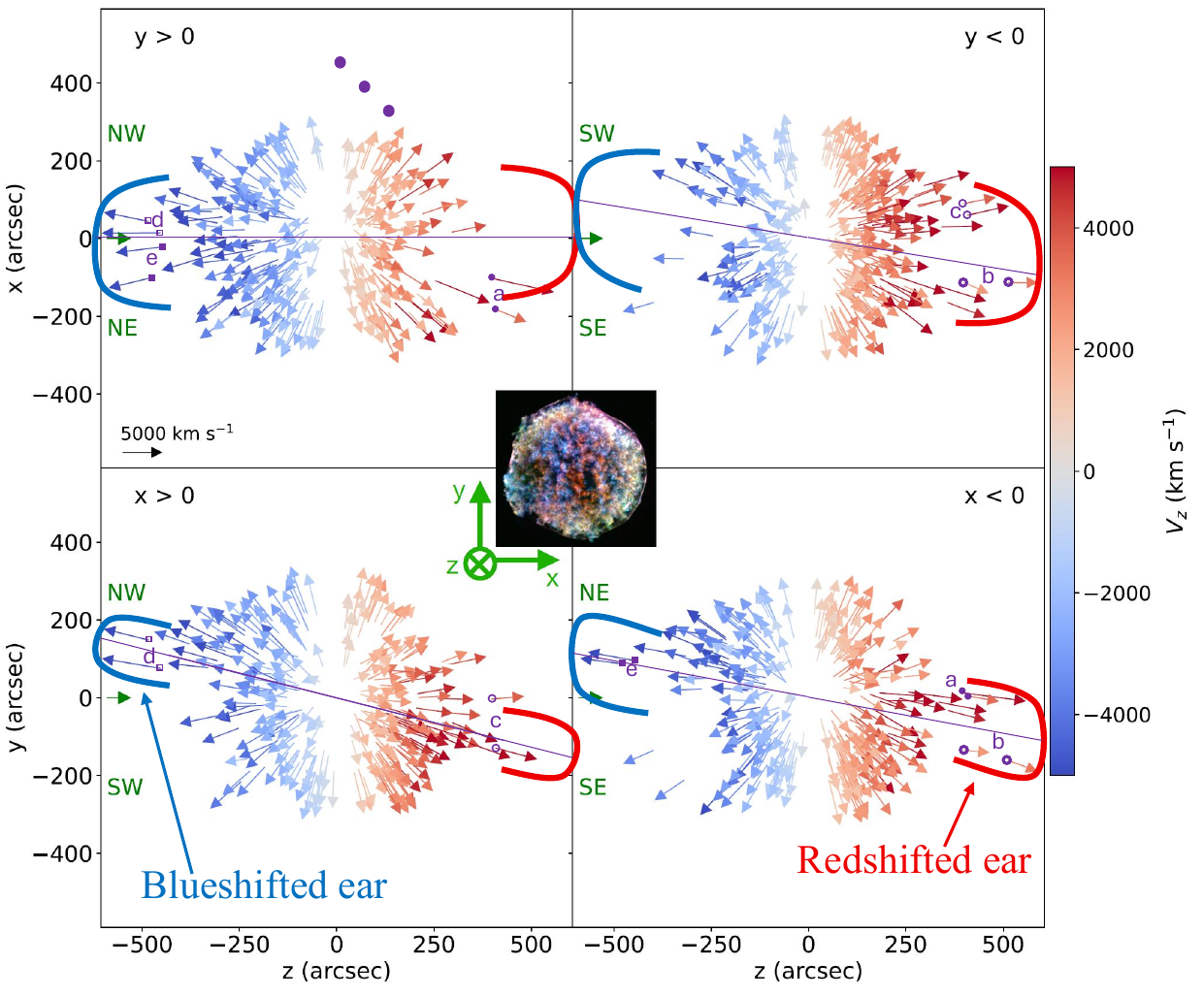} 
\caption{
An image adapted from \cite{Godinaudetal2023} showing the three-dimensional vector field of the velocities they deduce in their study. The velocity maps are of four half-spheres, as indicated in each of the four panels. 
The upper panels are views along the y-axis (from above). The bottom panels are views along the x-axis (from the right). The colors are the velocities along the line of sight, $V_z$. The green arrow at the left of each panel indicates the position of the observer. The lack of arrows near $z=0$ is due to a selection bias. 
\href{https://sketchfab.com/3d-models/3d-dynamics-of-tychos-supernova-remnant-5454760018c44c4da79c41e639179abb}{This is the link to the visualization of their 3D map.}
I added pairs of arcs that mark the general directions of the ears using thick blue and red lines, and identified 10 velocity arrows in five pairs. Pairs a-c belong to the redshifted ear, and pairs d and e to the blueshifted ear. In each panel, the two arcs are symmetric about the center, but these arcs do not represent actual ears, as the ears are smaller than the arcs (see text).  
}
\label{Fig:TychoGodinaud2023}
\end{center}
\end{figure*}

Figure \ref{Fig:TychoGodinaud2023}, which I adapted from  \cite{Godinaudetal2023}, presents their three-dimensional vector field of the velocities they deduced from their analysis. I notice some velocity arrows for ejecta knots farther from the center that have high velocities. I mark five pairs of ejecta knots in the outer shell: pairs a-c are redshifted, while pairs d and e are blueshifted. The redshifted knots a-c (three pairs of two knots) form a protrusion, the \textit{redshifted ear}, away from us and tilted slightly towards the south ($-y$ direction), while the blueshifted knots d and e form a protrusion towards us, the \textit{blueshifted ear}, tilted slightly to the north ($+y$). 
In each panel of Figure \ref{Fig:TychoGodinaud2023}, I mark the direction of the two ears in that pair with two arcs, a red arc for the redshifted ear and a blue arc for the blueshifted ear. These arcs mark the direction of the ears, not their actual size or structure. The ears are smaller than the arcs. 

I conclude that the three-dimensional structure of Tycho contains two opposite ears. Qualitatively, the pair of opposite ears is similar to the pairs in three other SNe Ia, namely, Kepler, SNR G299-2.9, and SNR G1.9+0.3. Previous studies suggested that each of these SNRs is an SN Ia exploding inside an old planetary nebula, i.e., they are SNIPs, and that the ears result from the pre-explosion planetary nebula structures:  \cite{TsebrenkoSoker2015} for Kepler and tentatively for SNR G299-2.9, and \cite{Soker2024RAAPN} for SNR G1.9+0.3.
I take the ears of Kepler SNR to be two opposite polar structures (e.g., \citealt{TsebrenkoSoker2013}), and not an equatorial protrusion as, e.g., \cite{Chiotellisetal2021} assume. 
I conclude that, as \cite{DickelJones1985} already suggested based on spherical analysis, Tycho is an SNIP, with the ears resulting from the pre-explosion structure of the old planetary nebula.  

\section{The possible scenarios for Tycho}
\label{sec:Scenarios}

In searching for the most appropriate SN Ia scenario for the Tycho SNR, I consider the following observations (see the reviews cited in Section \ref{sec:intro} for the different scenarios).
(1) Tycho SNR interacts with a massive CSM, $M_{\rm CSM} \simeq 1.3-1.7 M_\odot$ \citep{Kobashietal2024}. 
(2) There is no identified surviving companion star in Tycho (e.g., \citealt{RuizLapuenteetal2004, Kerzendorfetal2009, Kerzendorfetal2013, Kerzendorfetal2018, XueSchaefer2015}). In particular, there is no companion remnant of an asymptotic giant branch or red giant branch star, as required to form a massive CSM under the SD scenario.  
(3) Tycho is a normal SN Ia (e.g., \citealt{RuizLapuente2004, RuizLapuente2023, Badenesetal2006, Krauseetal2008}).

\textit{The WD-WD collision (WWC) scenario.} The presence of a massive CSM almost rules out the WWC scenario, where two unbound WDs collide and both explode. Only if one of the WDs happens to have an old planetary nebula around it could this scenario account for Tycho; this extremely rare event is still a SNIP. 

\textit{The singe degenerate (SD) scenario.} In the SD scenario, a giant mass-donor star is expected to form a massive CSM. However, the CSM mass of Tycho is too large for the SD scenario, and in any case, there is no surviving remnant of a giant star, which should be very bright at 500 years after explosion. 
More generally, the SD scenario appears to encounter additional severe challenges, for example, the WD mass does not increase in symbiotic systems as observations (e.g., \citealt{Schaefer2025, SchaeferMyers2025} for recent papers) and calculations (e.g., \citealt{Vathachiraetal2024, Vathachiraetal2025}) show; the SD scenario requires symbiotic progenitors (or a similar binary system with a giant) to supply a massive CSM as observed in the Tycho SNR. Moreover, if the WD does grow in mass, it will explode as a peculiar rather than a normal SN Ia (e.g., \citealt{MichaelisPerets2025})\footnote{Despite its clear failure to explain normal SNe Ia, many papers still mention the SD scenario as "one of the two scenarios for SNe Ia", indicative of the conservative nature of our society rather than the common SN Ia scenario in nature.\label{fnSD}} 

\textit{The double-degenerate (DD) and DD-MED scenarios.} In the DD scenario, the interaction of two bound WDs explodes both WDs on a dynamical timescale. In the DD-MED scenario, the two WDs merge, and an explosion occurs at a later time—the merger-to-explosion delay (MED)—which is much longer than the dynamical time scale; at explosion, there is a lonely WD. The DD or DD-MED scenarios are possible if the merger of the two WDs occurs within hundreds of thousands of years of the youngest WD's formation, such that its planetary nebula forms the CSM. Most studies attribute gravitational wave emission to triggering the merger, a very slow process. The probability of merger within $\simeq 10^6 \yr$ exists, but it is very low. These rare channels of the DD and DD-MED scenarios are also SNIPs. 

\textit{The double-detonation (DDet) and DDet34 scenarios.} In the DDet scenario, one WD accretes helium-rich gas from a companion and explodes. The companion survives. \cite{MandalSetal2025Tycho} analyze the sizes of small-scale turbulent substructures in Tycho and argue that it is most consistent with the DDet scenario. However, because no surviving companion has been detected in Tycho, this scenario is highly unlikely. If the mass-donor WD also explodes, then no remnant survives. This is possible if the shock wave of the first WD to explode, ignites the entire mass-donor WD, either in a third detonation, the triple-detonation channel, that explodes a helium WD (DDet3), or it ignites the helium outer shell, which then detonates the inner CO core of the second WD, namely, a total of four detonations, the quadrupole channel (DDet4); I mark them together DDet34 (see, e.g., \citealt{Pollinetal2025} for a recent paper on the DDet and DDet4 scenarios; note that they find that the DDet4 does not fully explain normal SNe Ia, and might better fit peculiar SNe Ia). 
In principle, the DDet34 that does not leave a surviving remnant can also explain the Tycho SNR. The CSM, then, comes from the planetary nebula of the younger WD. Again, gravitational waves should cause the two WDs to approach each other, making this scenario with a massive CSM very rare. This scenario is also an SN Ia inside an old planetary nebula, i.e., an SNIP. 

\textit{The core-detonation (CD) scenario.} In the CD scenario, a WD spirals inside the envelope of an AGB star and merges with the core. The explosion occurs much later, namely, the MED time (the time from merger to explosion) is much longer than the dynamical time.  
In the CD scenario and the DD-MED scenario, at the time of explosion and within several dynamical times before the explosion, there is only one WD, no companion with which it interacts. This is a lonely WD; lonely and not single, because the lonely CO WD is a merger product of a binary system, and not a product of single-stellar evolution. 

I claimed (e.g., \citealt{Soker2024Rev}) that most, or even all, normal SNe Ia are descendants of lonely WD scenarios (the CD and DD-MED scenarios), while peculiar SNe Ia might result from all scenarios. 
The lonely WD does not survive the explosion in normal SNe Ia, and might survive in some peculiar SNe Ia. The MED time should be much longer than the dynamical time of the merger process, by up to more than thirteen orders of magnitude (amounts to $\simeq 10^6 \yr$). This allows the WD remnant of the merger to relax and form a large-scale spherical explosion. The notion of MED time in SN Ia scenarios is several years old (e.g., \citealt{Soker2018Rev, Soker2022RAA, Neopaneetal2022}), while the idea of the group of lonely WD scenarios is two years old \citep{Soker2024Rev}.

Studies attributed several SNe Ia (e.g., PTF 11kx, \citealt{Sokeretal2013}; SN LSQ14fmg, \citealt{Hsiaoetal2020}; 2003fg-like [``super-Chandrasekhar''] SNe Ia, \citealt{Ashalletal2021}; the 03fg-like SN Ia ASASSN-15hy, \citealt{Luetal2021}; SN 2020uem, \citealt{Unoetal2023}) and SN Ia remnants (e.g., SNR G1.9+0.3, \citealt{Soker2024RAAPN}; SNR 0509-67.5, \citealt{Soker2025SNR0509}) to the CD scenario. 
Other studies support the CD scenario by examining common properties of several SNe Ia, such as polarization (e.g., \citealt{Cikotaetal2017}). 
In a recent study \citep{Soker2025DDMED}, I suggested that SN Ia SN 2020aeuh (for its peculiar carbon-oxygen-rich CSM see \citealt{Tsalapatasetal2025}), is a descendant of the DD-MED scenario with a MED time of $\simeq 1-2 \yr$.  

The CD scenario includes a MED time, and it is likely to explode while the planetary nebula shell is still intact.  Therefore, the CD scenario is the most likely scenario for Tycho. Less likely, but possible, are the DD, DD-MED, and the DDet34 scenarios. In all cases, the massive CSM implies an explosion inside a planetary nebula, i.e., an SNIP. (The WWC has a very slight probability of being the relevant scenario for Tycho.)
The end of the common envelope evolution of the Tycho progenitor system is either a core-WD merger in the CD scenario, or a very tight orbital separation that allows merger by gravitational waves within several hundred thousand years, so that the planetary nebula is still around, in the DD, DD-MED, and DDet34 scenarios. 
This, in turn, requires the envelope through which the WD spirals in to be massive to bring it close to the core. 

\section{Implications}
\label{sec:Implication}

\subsection{The fraction of SNIPs among normal SNe Ia}
\label{subsec:SNIPsFraction}

I first note that the class of SNe Ia-CSM is distinct from that of SNIPs. SNe Ia-CSM (e.g., \citealt{Sharmaetal2023, Terweletal2025a, Terweletal2025b}, and references to earlier papers therein)  are SNe Ia that interact with a CSM within hours to several years of explosion. The CSM might be of small mass and not of planetary nebula origin, or might be a young planetary nebula. The class of SNe Ia-CSM is much-much smaller than SNIPs, e.g., \cite{Sharmaetal2023} estimated the fraction of SNe Ia-CSM to be $\simeq 0.02 -0.2 \%$. In the majority of SNIPs, the ejecta-CSM interaction occurs much later after the explosion (e.g., \citealt{Courtetal2024} for a theoretical study).

\cite{TsebrenkoSoker2015} estimated the fraction of SNIPs among all normal SNe Ia to be $f_{\rm SNIP}(2015) \approx 0.1-0.3$. In \cite{Soker2019CEEDTD} I still estimated the fraction of SNIP as $f_{\rm SNIP} (2019) \simeq 0.15 - 0.20$. 
In \cite{Soker2022RAA} I raised the value to $f_{\rm SNIP} (2022) \simeq 0.5$. This large increase in the estimated value of $f_{\rm SNIP}$ came mainly from the work of \cite{Lietal2021} who found massive CSM in some SNRs Ia in the Large Magellanic Cloud (LMC). 
In \cite{Soker2022RAA} I estimated the fraction of SNIPs in the MW and the LMC to be $f_{\rm SNIP}(2022)_{\rm (MW+LMC)} \simeq 0.7-0.8$. I considered in that study the selection effects that make SNIPs brighter for longer times because the massive CSM slows their expansion and makes them brighter. I also considered that the two Galaxies have ongoing star formation. For these selection effects I estimated the value $f_{\rm SNIP} (2022) \simeq 0.5 < f_{\rm SNIP}(2022)_{\rm (MW+LMC)}$. 
In \cite{Soker2022RAA}, the SNR 0509-67.5 was not listed as an SNIP. In \cite{Soker2025SNR0509} I argued that SNR 0509-67.5 is an SNIP and raised the estimate to over $50\%$,  $f_{\rm SNIP} (2025) > 0.5$.

In \cite{Soker2022RAA}, I listed Tycho as `maybe' an SNIP because of its Balmer-dominated spectrum \citep{KirshnerChevalier1978}, which might indicate a massive CSM. In this study, I changed its status to a secure SNIP. Going back to the calculation in \cite{Soker2022RAA}, out of 15 SNRs Ia, 13 are SNIPs, and two are `maybe' SNIPS. This brings me to estimate the SNIP fraction in the two galaxies to be  
\begin{equation}
f_{\rm SNIP}(2026)_{\rm (MW+LMC)} \simeq 0.87-1. 
\label{eq:fSNIPMW}
\end{equation}
The upper limit of $100 \%$ of normal SNRs Ia in the MW and LMC raises the possibility that there are no selection effects, and that all normal SNe Ia are SNIPs. It is more likely that some normal SNe Ia are from the DD-MED scenario. The merger might take a very long time after the formation of the double WD system. I therefore crudely take the SNIP fraction among normal SNe Ia in the local Universe, including galaxies with no ongoing star formation, to be  
\begin{equation}
f_{\rm SNIP}(2026) \simeq 0.7-0.9. 
\label{eq:fSNIPtotal}
\end{equation}

In \cite{Soker2025SNR0509}, I suggested that SNR~0509-67.5 is an SNIP, and from that deduced $f_{\rm SNIP} (2025) > 0.5$. I noticed that the large fraction of SNIPs I deduced is compatible with the observational finding by \cite{Moetal2025} of a larger estimated number of SNe Ia experiencing late interaction with a CSM. 
The new estimate in equation (\ref{eq:fSNIPtotal}) strengthens the earlier conclusion. I now clearly state my new conclusion that \textit{the vast majority of normal SNe Ia are SNIPs.} 

\subsection{The delay time distribution (DTD)}
\label{subsec:DTD}

Different groups (e.g., \citealt{Grauretal2014, Heringeretal2017, MaozGraur2017, Frohmaieretal2019, Wisemanetal2021, Toyetal2023, Joshietal2024, Palicioetal2024}) infer distinct delay-time distributions (DTDs). The DTD is the distribution of SNe Ia as a function of time after a burst of star formation.  In \cite{Soker2022RAA} I used DTDs from \cite{FriedmannMaoz2018} and \cite{Heringeretal2019} to derive a DTD of the form  
\begin{equation}
\begin{aligned}
\dot N_{\rm DTD} &  =  0.147 N_{\rm Ia} \left( \frac{t_{\rm SF-E}}{1 \Gyr} \right)^{-1.32} \Gyr^{-1} ,
\\ &
 {\rm for} \quad 0.05 \Gyr < t_{\rm SF-E} < 13.7 \Gyr, 
\label{eq:DTD}
\end{aligned}
\end{equation}
where $t_{\rm SF-E}$ is the time from star formation to the explosion and $N_{\rm Ia}$ is the total number of normal SN Ia in the sample.  The minimum time of $0.05 \Gyr$ in equation (\ref{eq:DTD}) for the formation of an SN Ia after star formation corresponds to a main-sequence mass of $M_{\rm ZAMS} \simeq 6 M_\odot$. 
I found in \cite{Soker2022RAA} that if all SNe Ia with $0.05 \Gyr < t _{\rm SF-E} < 0.27 \Gyr$ explode as SNIPS, then $f_{\rm SNIP} (2022) \simeq 0.5$. This time range corresponds to progenitors of an initial mass range of $3.5 M_\odot \la M_{\rm ZAMS} \la 6 M_\odot$. 

To accommodate the larger SNIP fraction I find here, the time range and the corresponding approximate mass range should be $0.05 \Gyr < t _{\rm SF-E} < 0.77 \Gyr$ and $2.5  M_\odot \la M_{\rm ZAMS} \la 6 M_\odot$ for $f_{\rm SNIP} = 0.7$,  and $0.05 \Gyr < t _{\rm SF-E} < 1.56 \Gyr$ and $2 M_\odot \la M_{\rm ZAMS} \la 6 M_\odot$ for $f_{\rm SNIP} = 0.8$. 

A key point is that the second star to form an AGB star must have a massive enough envelope to force the older WD to spiral all the way and merge with the core (or come very close to it, such that gravitational waves lead to merger within several hundred years). There is no determined mass limit; a crude guess is $M_2 \gtrsim 3 M_\odot$ (e.g., \citealt{Sokeretal2013}). The zero-age main-sequence mass might be smaller, as the secondary star might accrete mass from the primary.  
I have the following points to make to strengthen my conclusion from \cite{Soker2025SNR0509} regarding the DTD. 
\cite{Badenesetal2009} argued that SNR~0509-67.5 resides in a population with a mean age of $\tau_{\rm SF} \simeq 7.9 \times 10^9 \yr$, but a fraction of the population might be younger with $\tau_{\rm SF} \lesssim 1.8 \times 10^8 \yr$. The latter time corresponds to stars with $M_{2,{\rm ZAMS}} \simeq 4-6 M_\odot$ that now form WDs. I suggest that SNR~0509--67.5 belongs to the younger population.   
I speculated in \cite{Soker2025SNR0509} that many SNe Ia in old stellar populations belong to younger subpopulations. Namely, that these `dead galaxies' have pockets of more recent star formation.  

In this study, with the larger fraction of SNIPs (equation \ref{eq:fSNIPtotal}), I strengthen the speculation from \cite{Soker2025SNR0509} and strongly argue that the DTD commonly attributed solely to the stellar evolution from star formation to the SN Ia explosion \textit{also includes delay pockets associated with late star formation in galaxies. }

\section{Summary}
\label{sec:Summary}

In examining recent observational studies of SNR Ia Tycho, I concluded that \textit{Tycho exploded inside a planetary nebula} (or a shell of a planetary nebula remnant, which is no longer ionized), namely, an SNIP (Section \ref{sec:Geometry}).  I put the 40-year-old suggestion by \cite{DickelJones1985} that Tycho exploded inside a planetary nebula on solid ground. They used hydrodynamical calculations in spherical symmetry; I consider the non-spherical morphology. 
In the morphology of the ejecta that  \cite{Godinaudetal2023} inferred, I identified two ears. I marked the directions of the two ears in this pair of ears (but not their size) in Figure \ref{Fig:TychoGodinaud2023}. The pair of Tycho's ears I identified resembles those of three other SNRs Ia that earlier studies have attributed to SNIPs (Kepler, SNR G299-2.9, and SNR G1.9+0.3).

In Section \ref{sec:Scenarios}, I discuss different possible scenarios for Tycho. I ruled out the DDet and SD scenarios for Tycho because there is no surviving companion. I also note that the SD is unlikely to explain any normal SN Ia (see footnote \ref{fnSD} on page \pageref{fnSD}). The WWC scenario is highly unlikely, while the DD, DD-MED, and DDet34 are possible, but unlikely, for Tycho. \textit{I find the CD scenario to be the most likely one for Tycho.} All the scenarios, likely, unlikely, or highly unlikely, must be SNIPs. 

Adding Tycho to the list of SNIPs, I estimated (Section \ref{subsec:SNIPsFraction}) the fraction of SNIPs among all normal SNe Ia in the MW and LMC to be $f_{\rm SNIP}(2026)_{\rm (MW+LMC)} \simeq 0.87-1$ (equation \ref{eq:fSNIPMW}). I then crudely estimated the fraction of SNIPs among all SNe Ia, including in galaxies without ongoing star formation, to be  $f_{\rm SNIP}(2026) \simeq 0.7-0.9$ (equation \ref{eq:fSNIPtotal}). This fraction is highly uncertain, but sufficient to bring me to firmly state my view that \textit{the vast majority of normal SNe Ia are SNIPs.}

In Section \ref{subsec:DTD}, I discussed the implications of the newly determined large value of $f_{\rm SNIP}$ on the DTD of normal SNe Ia. To accommodate the large fraction of SNIPs and without recent star formation in old stellar populations, binary systems with a secondary mass as low as $M_{\rm ZAMS} \simeq 2 M_\odot$ should form normal SNe Ia. However, it is not clear whether such stars have an envelope sufficiently massive during their AGB phase to force the older WD to merge with the core in the CD scenario. Instead, I considered that galaxies with an old stellar population had pockets of more recent star formation. Namely, normal SNe Ia come from younger stellar populations, with ages of $\lesssim 0.5-1 \Gyr$, rather than from very old populations. This conclusion is compatible with the very recent finding by \cite{Scherbaketal2025} that the Ca-rich gap transients and 91bg-like SNe, which are faint peculiar SNe Ia, have the longest peak delay times in their DTD,  $\approx 10 \Gyr$. In contrast, the peak delay time of SNe Ia DTD is only $\approx 1 \Gyr$ (of Type II SNe it is $\approx 10 \Myr$). 
I concluded Section \ref{subsec:DTD} by putting on a more solid ground an earlier speculation \citep{Soker2025SNR0509} that \textit{the DTD of normal SNe Ia includes not only the stellar evolution timescale from the star formation epoch of the main stellar population to SN Ia explosion (as usually assumed), but also includes pockets of stellar populations associated with later star formation than the average in quiescent galaxies } (those without ongoing star formation).

More generally, this study supports the suggestion that normal SNe Ia result from lonely WDs, namely, massive WDs that are remnants of either a two-WD merger in the DD-MED scenario or a core-WD merger in the CD scenario. 

\section*{Acknowledgments}
I thank Gunter Cibis for helpful comments. I thank the Charles Wolfson Academic Chair at the Technion for the support.





\end{document}